\documentclass[12pt]{article}
\usepackage{amssymb}
\usepackage{amsmath,amsfonts,feynmp}
\usepackage{tipa}
\usepackage{stmaryrd}

\makeatletter\@addtoreset{equation}{section}\makeatother

\def\be{\begin{equation}}
\def\ee{\end{equation}}
\def\bea{\begin{eqnarray}}
\def\eea{\end{eqnarray}}

\makeatletter\@addtoreset{equation}{section}\makeatother

\renewcommand{\title}[1]{\vbox{\center\LARGE{#1}}\vspace{5mm}}
\renewcommand{\author}[1]{\vbox{\center#1}\vspace{5mm}}
\newcommand{\address}[1]{\vbox{\center\em#1}}

\begin{document}

\unitlength = .8mm

\begin{titlepage}
\begin{center}
\hfill \\
\hfill \\
\vskip 1.5cm

\title{On the Generalized Massive Gravity in $AdS_3$ }

\vskip 0.5cm
 {Yan Liu\footnote{Email: liuyan@itp.ac.cn}} and {Ya-Wen
Sun\footnote{Email: sunyw@itp.ac.cn}}

\address{ Key Laboratory of Frontiers in Theoretical Physics£¬
\\ Institute of Theoretical Physics, Chinese Academy of Sciences
\\P.O. Box 2735, Beijing 100190, China}

\end{center}

\vskip 1cm

\abstract{In this note we investigate the generalized massive
gravity in asymptotically $AdS_3$ spacetime by combining the two
mass terms of topological massive gravity and new massive gravity
theory. We study the linearized excitations around the $AdS_3$
background and find that at a specific value of a certain
combination of the two mass parameters (chiral line), one of the
massive graviton solutions becomes the left moving massless mode. It
is shown that the theory is chiral at this line under Brown-Henneaux
boundary condition. Because of this degeneration of the gravitons
the new log solution which has a logarithmic asymptotic behavior is
also a solution to this gravity theory at the chiral line. The log
boundary condition which was proposed to accommodate this log
solution is proved to be consistent at this chiral line. The
resulting theory is no longer chiral except at a special point on
the chiral line, where another new solution with log-square
asymptotic behavior exists. At this special point, we prove that a
new kind of boundary condition called log-square boundary condition,
which accommodates this new solution, can be consistent. }

\vfill

\end{titlepage}

\eject \tableofcontents

\section{Introduction}

The anti-de Sitter /conformal field theory (AdS/CFT) correspondence
\cite{Maldacena:1997re,{Gubser:1998bc},{Witten:1998qj},{Aharony:1999ti}}
gives much insight into the quantum theory of gravity and the
central idea is that what it means to solve the quantum gravity with
a negative cosmological constant is to find the dual conformal field
theory. It would be interesting to test this idea in three
dimensional pure quantum gravity with a negative cosmological
constant and find as much useful information of the dual conformal
field theory from the gravity side. The central charge of the
corresponding conformal field theory has been derived in
\cite{Brown:1986nw} two decades ago from the observation that the
asymptotic symmetry group of $AdS_3$ under Brown-Henneaux boundary
conditions possesses two sets of Virasoro algebras and the Hilbert
space should be a representation of this algebra. Much work has been
done on $AdS_3/CFT_2$ to gain valuable insights into the
quantization of gravity in asymptotically $AdS_3$, which can be
found in \cite{Carlip:2005zn,{Kraus:2006wn}} and references therein.

Over the past two years, two remarkable progresses have been made in
$AdS_3/CFT_2$. One is for pure gravity in $AdS_3$ and the CFT dual
of this quantum gravity has been identified in
\cite{{Witten:1988hc},Witten:2007kt}. The other progress is that the
field theory dual for a Chern-Simons (CS) deformation of pure
Einstein gravity theory named topological massive gravity (TMG)
\cite{Deser:1981wh,{Deser:1982vy}} with a negative cosmological
constant has been investigated and it is found that there exists a
chiral point at which the theory becomes chiral with only
right-moving modes \cite{Li:2008dq}. TMG has been further discussed
in \cite{Carlip:2008jk,{Hotta:2008yq},{Grumiller:2008qz},
{Li:2008yz},{Park:2008yy},{Sachs:2008gt},{Grumiller:2008pr},{Carlip:2008eq},
{Lee:2008gta},{Sachs:2008yi},{Gibbons:2008vi},{Anninos:2008fx},{Carlip:2008qh},{Giribet:2008bw},{Strominger:2008dp},
{Compere:2008cv},{Myung:2008dm},{Grumiller:2008es},{Garbarz:2008qn},{Blagojevic:2008bn},
{Henneaux:2009pw},{Sezgin:2009dj},{Maloney:2009ck},{Oh:2009if}}.

In \cite{Bergshoeff:2009hq} another interesting massive deformation
of pure gravity has been proposed in three dimensions. In this new
massive gravity (NMG), higher derivative terms are added to the
Einstein Hilbert action and unlike in topological massive gravity,
parity is
 preserved in this new massive gravity. This new massive gravity is
 equivalent to the Pauli-Fierz action for a massive spin-2 field at
the linearized level in asymptotically Minkowski spacetime. In
\cite{Nakasone:2009bn, {Nakasone:2009vt}}, the unitarity of this new
massive gravity and the new massive gravity with a Pauli-Fierz mass
term was examined. Black hole solutions for this new massive gravity
with a negative cosmological constant have been considered in
\cite{Clement:2009gq}. In \cite{Liu:2009bk}, the linearized
gravitational excitations of this new massive gravity around
asymptotically $AdS_3$ spacetime has been studied and it was also
found that there is a critical point for the mass parameter. It was
conjectured in \cite{Liu:2009bk} that this theory is trivial at the
critical point under Brown-Henneaux boundary condition, and was
later proved in \cite{Liu:2009kc}.  Other consistent boundary
conditions are also studied in \cite{Liu:2009kc}.

TMG and NMG have much in common, and indeed can be unified into a
General Massive Gravity theory \cite{Bergshoeff:2009hq} (GMG). The
generalized massive gravity theory is realized by adding both the CS
deformation term and the higher derivative deformation term to pure
Einstein gravity with a negative cosmological constant. This theory
has two mass parameters and TMG and NMG are just two different
limits of this generalized theory. This theory is expected to have
more interesting physics because we can have one more adjustable
mass parameter.

In this note we will study this generalized massive gravity theory
in $AdS_3$. We begin with investigations on the behavior of
linearized excitations of GMG in asymptotically $AdS_3$. We show
that there exists a critical line of the two mass parameters at
which one of the massive graviton becomes massless like what happens
at the chiral point in TMG. At this critical line the new solution
with log asymptotic behavior which was found in
\cite{Grumiller:2008qz} for TMG is also a solution to this GMG. The
linearized behavior in this theory is different from that of TMG in
that there also exists a special point on the critical line at which
both of the two massive gravitons become left-moving massless
gravitons and new interesting physics arises. This special point is
also different from in NMG, where parity is preserved and at the
special point the two massive gravitons become one left-moving
massless and one right-moving massless modes. At this special point
in GMG, another new solution which has log-square asymptotic
behavior appears.

 The GMG theory at the critical line has zero left-moving central
 charge and is shown to
be chiral under Brown-Henneaux boundary condition at the linearized
level. Because of the appearance of the new log solution, we try to
relax the boundary condition to the log boundary condition and find
that this boundary condition is consistent with GMG at the chiral
line. The theory is no longer chiral under the log boundary
condition except at the special point mentioned above. At the
special point, with the hint from the asymptotic behavior of the new
solution which does not obey either the Brown-Henneaux boundary
condition or the log boundary condition, we can further relax the
log boundary condition to the new log-square boundary condition and
find it is indeed consistent at this point.

Our note is organized as follows. In Sec.2 we will formulate the
generalized massive gravity in asymptotically $AdS_3$ and consider
the linearized excitations. In Sec.3 we will first write out the
formula for the calculation of conserved charges in this theory.
Then we will examine the consistency of the Brown-Henneaux boundary
condition, the Log boundary condition at the chiral line and the
log-square boundary condition at the special point. Sec.4 is devoted
to conclusions and discussions.

\section{The Generalized Massive Gravity Theory}

In this section we will first write out the action of the
generalized massive gravity theory which is obtained by combining
the two mass terms of TMG and NMG. We will show that the theory can
be treated as a generalized massive gravity in asymptotically
$AdS_3$ in the sense that there are two independent mass parameters
in the linearized equation of motion after gauge fixing. This is
similar to the case without a cosmological constant discussed in
\cite{Bergshoeff:2009hq}. Then we will study the linearized
solutions of this generalized massive gravity around $AdS_3$.

\subsection{The Generalized Massive Gravity Theory}
The action for the generalized massive gravity theory
\cite{Bergshoeff:2009hq} can be written as\footnote{ We take the
metric signature(-,+,+) and follow the notation and conventions of
MTW \cite{Misner:1974qy}.
 $G$ is the three dimensional Newton constant which is
positive here. We take
$\varepsilon_{\mu\nu\alpha}=\sqrt{-g}\epsilon_{\mu\nu\alpha}$ with
$\epsilon_{012}=-1$. Our convention is the same as
\cite{Li:2008dq}.}

 \be I=\frac{1}{16\pi
G}\int d^3x \sqrt{-g} \bigg[(R-2\lambda)+\frac{1}{\mu}{\mathcal
L}_{CS}-\frac{1}{m^2}K\bigg], \ee where \be
K=R^{\mu\nu}R_{\mu\nu}-\frac{3}{8}R^2,\ee \be {\mathcal
L}_{CS}=\frac{1}{2}\varepsilon^{\lambda\mu\nu}\Gamma^{\alpha}_{\lambda\sigma}
[\partial_{\mu}\Gamma^{\sigma}_{\alpha\nu}+\frac{2}{3}\Gamma^{\sigma}_{\mu\tau}\Gamma^{\tau}_{\nu\alpha}],\ee
$m$, $\mu$ are the two mass parameters of this generalized massive
gravity and $\lambda$ is a constant which is different from the
cosmological constant of the $AdS_3$ background solution. The sign
of $m^2$ is not fixed here and it can be either positive or
negative. We assume $\mu$ to be positive and cases with negative
$\mu$ can be obtained by exchanging the coordinates $\tau^+$ and
$\tau^-$, whose definition can be found below.

The equation of motion of this action is \be G_{\mu\nu}+\lambda
g_{\mu\nu}-\frac{1}{2m^2}K_{\mu\nu}+\frac{1}{\mu}C_{\mu\nu}=0
\label{eom},\ee where \be
G_{\mu\nu}=R_{\mu\nu}-\frac{1}{2}g_{\mu\nu}R,\ee \bea
K_{\mu\nu}&=&-\frac{1}{2}\nabla^2 R
g_{\mu\nu}-\frac{1}{2}\nabla_\mu\nabla_\nu R+2\nabla^2 R_{\mu\nu}
\nonumber\\&&+4R_{\mu \alpha \nu \beta}R^{\alpha
\beta}-\frac{3}{2}RR_{\mu\nu}-R_{\alpha\beta}R^{\alpha\beta}g_{\mu\nu}+\frac{3}{8}R^2g_{\mu\nu},
\eea and the Cotton tensor \be
C_{\mu\nu}=\varepsilon_{\mu}^{~\alpha\beta}\nabla_{\alpha}(R_{\beta\nu}-\frac{1}{4}g_{\beta\nu}R).\ee
One special feature of this choice of $K$ is that $g^{\mu\nu}
K_{\mu\nu}=K.$

After introducing a non-zero $\lambda$, the new massive gravity
theory could have an $AdS_3$ solution
\begin{equation}\label{AdS3metric}
ds^2=\bar{g}_{\mu\nu}dx^\mu dx^\nu=\ell^2(-
\cosh^2{\rho}d\tau^2+\sinh^2{\rho}d\phi^2+d\rho^2),
\end{equation}
while the $\lambda$ in the action should be related to the
cosmological constant $\Lambda$ and the mass parameter by
 \be
m^2=\frac{\Lambda^2}{4(-\lambda+\Lambda)},\ee and \be
\Lambda=-1/\ell^2.\ee Note that for a given $\lambda$, there may
exist two $AdS_3$ solutions \cite{Bergshoeff:2009hq} with different
AdS radii. We can expand around either one of them and denote $\ell$
as the AdS radius of the one we expand around. The other solution
has different asymptotic behavior and will not affect our result.

It would be useful to introduce the light-cone coordinates
$\tau^{\pm}=\tau \pm \phi$, then the $AdS_3$ spacetime
(\ref{AdS3metric}) could be written as
\begin{equation}
ds^2=\frac{\ell^2}{4}(-d\tau^{+2}-2
\cosh{2\rho}d\tau^{+}d\tau^{-}-d\tau^{-2}+4d\rho^2).
\end{equation}

The central charges
\cite{{Sahoo:2006vz},{Park:2006gt},Park:2006zw,{Kraus:2006wn}} for
this theory in asymptotically $AdS_3$ are
 \be
c_L=\frac{3\ell}{2G}(1-\frac{1}{2m^2\ell^2}-\frac{1}{\mu \ell}),~~~~
c_R=\frac{3\ell}{2G}(1-\frac{1}{2m^2\ell^2}+\frac{1}{\mu \ell}). \ee
To have nonnegative central charges we have to impose the
constraints $c_L\geq0$ and $c_R\geq0$ on the mass parameters.

\subsection{Linearized Equation of Motion}

By expanding $g_{\mu\nu}=\bar{g}_{\mu\nu}+h_{\mu\nu}$ around the
$AdS_3$ background solution (\ref{AdS3metric}), we could obtain the
equation of motion for the linearized excitations $h_{\mu\nu}$ as
\be\label{eomold} G_{\mu\nu}^{(1)}+\lambda
h_{\mu\nu}+\frac{1}{\mu}C_{\mu\nu}^{(1)}-\frac{1}{2m^2}K_{\mu\nu}^{(1)}=0,\ee
where
\begin{eqnarray} R_{\mu\nu}^{(1)}&
=& \frac{1}{2} (- \bar{\nabla}^2  {h}_{\mu\nu} - \bar{\nabla}_{\mu}
\bar{\nabla}_{\nu}  h + \bar{\nabla}^{\sigma} \bar{\nabla}_{\nu}
 h_{\sigma\mu} + \bar{\nabla}^{\sigma}  \bar{\nabla}_{\mu}  h_{\sigma\nu}),\\
R^{(1)} &\equiv& (R_{\mu\nu}  g^{\mu\nu})^{(1)} = - \bar{\nabla}^2 h
+ \bar{\nabla}_{\mu} \bar{\nabla}_{\nu}  h^{\mu\nu} - 2 \Lambda h,\\
G^{(1)}_{\mu\nu}&=& R^{(1)}_{\mu\nu}-\frac{1}{2}\bar{g}_{\mu\nu}R^{(1)}-3\Lambda h_{\mu\nu},  \\
K^{(1)}_{\mu\nu} &=& -\frac{1}{2}\bar{\nabla^2} R^{(1)}
\bar{g}_{\mu\nu}-\frac{1}{2}\bar{\nabla}_\mu\bar{\nabla}_\nu
R^{(1)}+2\bar{\nabla^2} R^{(1)}_{\mu\nu}-4\Lambda
\bar{\nabla}^2h_{\mu\nu}\nonumber\\&&-5\Lambda R^{(1)}_{\mu\nu}
+\frac{3}{2}\Lambda R^{(1)}\bar{g}_{\mu\nu}+\frac{19}{2}\Lambda^2
h_{\mu\nu},\\
C_{\mu\nu}^{(1)}&=&\varepsilon_{\mu}^{~\alpha\beta}\bar{\nabla}_{\alpha}
(R_{\beta\nu}^{(1)}-\frac{1}{4}\bar{g}_{\beta\nu}R^{(1)}-2\Lambda
h_{\beta\nu}).
\end{eqnarray}
We can also have $R^{(1)}=0$ by tracing both sides of the equation
of motion of gravitons (\ref{eomold}). Thus we can choose the gauge
to be $h=0$ and $\nabla_{\mu}h^{\mu\nu}=0$
simultaneously.\footnote{In fact, we can not have $R^{(1)}=0$ at the
point $m^2\ell^2=-1/2$, but we do not need to worry about this
because it will be shown in the next subsection that this point is
excluded from the parameter region that we are interested in.} After
gauge fixing, the equation of motion becomes
\be\label{eomforgraviton}
(\bar{\nabla^2}-2\Lambda)(\bar{\nabla^2}h_{\mu\nu}-\frac{m^2}{\mu}
\varepsilon_{\mu}^{~\alpha\beta}\bar{\nabla}_{\alpha}h_{\beta\nu}-(m^2+\frac{5}{2}\Lambda)h_{\mu\nu})=0.\ee
We can define the following four operators
\cite{Li:2008dq,{Bergshoeff:2009hq}} which commute with each other
as
\be(\mathcal{D}^{L/R})_\mu^{~\nu}=\delta_\mu^\nu\pm\ell\varepsilon_\mu^{~\alpha\nu}\bar{\nabla}_\alpha,\ee
\be
(\mathcal{D}^{m_i})_\mu^{~\nu}=\delta_\mu^\nu+\frac{1}{m_i}\varepsilon_\mu^{~\alpha\nu}\bar{\nabla}_\alpha,~i=1,2,\ee
and the linearized equation of motion can be written using these
four operators to be
\be\label{eee}(\mathcal{D}^{L}\mathcal{D}^{R}\mathcal{D}^{m_1}\mathcal{D}^{m_2}h)_{\mu\nu}=0,
\ee with $m_1m_2=-m^2-\frac{1}{2\ell^2}$ and
$m_1+m_2=-\frac{m^2}{\mu}$. At this linearized level, the
generalized massive gravity has two mass parameters $m_1$ and $m_2$
which can be different from each other. The value of $1/\mu$
reflects the extent to which parity is violated in this generalized
massive gravity. When $\mu\rightarrow\infty$, $m_1=-m_2$ and the
linearized equation of motion (\ref{eee}) goes back to the case of
NMG with parity preserved. When $m^2\rightarrow\pm\infty$, one of
$m_i$ goes to infinity and the other is equal to $\mu$, so one of
the operator $\mathcal{D}^{m_i}$ becomes the identity operator and
the linearized equation of motion (\ref{eee}) goes back to the case
of TMG. Thus we can see that TMG is a special case of this
generalized massive gravity in that one of the mass parameter goes
to infinity and NMG is a special case of this generalized massive
gravity in that the two mass parameters are equal in value while
with different signs. This gravity theory can be viewed as a
generalized massive gravity in the sense that the two mass
parameters $m_1$ and $m_2$ in the linearized equation of motion can
be independent with each other.

Therefore, because this generalized massive gravity is a more
generalized theory than TMG and NMG, it would be interesting to
investigate the consistent boundary conditions for this theory and
in the remainder of this paper we will focus on this generalized
massive gravity and found that there are other special points where
novel properties arise that the ones in TMG and NMG because $m_1$
and $m_2$ here can be chosen arbitrarily.

\subsection{Linearized Solutions}
We solve for the highest weight states which obey
$L_{0}|\psi_{\mu\nu}\rangle=h|\psi_{\mu\nu}\rangle$ and
$\bar{L}_{0}|\psi_{\mu\nu}\rangle=\bar{h}|\psi_{\mu\nu}\rangle$. We
can
have the following four sets of solutions \bea (h,\bar{h})&=&(2,0),\nonumber\\
(h,\bar{h})&=&(0,2), \nonumber\\
(h,\bar{h})&=&\Big(\frac{6-\frac{m^2\ell}{\mu}\pm\sqrt{2+4m^2\ell^2+\frac{m^4\ell^2}{\mu^2}}}{4},~~
\frac{-2-\frac{m^2\ell}{\mu}\pm\sqrt{2+4m^2\ell^2+\frac{m^4\ell^2}{\mu^2}}}{4}\Big),\nonumber\\
 (h,\bar{h})&=&\Big(\frac{-2+\frac{m^2\ell}{\mu}\pm
\sqrt{2+4m^2\ell^2+\frac{m^4\ell^2}{\mu^2}}}{4},~~\frac{6+\frac{m^2\ell}{\mu}\pm
\sqrt{2+4m^2\ell^2+\frac{m^4\ell^2}{\mu^2}}}{4}\Big),\nonumber\eea
where
\begin{eqnarray}&&\psi_{\mu\nu}(h,\bar{h})=f(\rho,\tau^+,\tau^-)\left(\begin{array}{ccc}
                                         1 & {h-\bar{h}\over2}& {i\over\sinh(\rho)\cosh(\rho)} \\
                                          {h-\bar{h}\over2} & 1 & {i(h-\bar{h})\over2\sinh(\rho)\cosh(\rho)} \\
                                         {i\over\sinh(\rho)\cosh(\rho)} &{i(h-\bar{h})\over2\sinh(\rho)\cosh(\rho)}   & - {1\over\sinh^2(\rho)\cosh^2(\rho)} \\
                                       \end{array}\right),\end{eqnarray}
                                       and $f(\rho,\tau^{+},\tau^{-})=e^{-ih\tau^{+}-i\bar{h}\tau^{-}}(\cosh
\rho)^{(-h-\bar{h})}\sinh^2\rho$.

The first two solutions are the left and right moving massless
gravitons respectively, which are solutions of
$(\mathcal{D}^{L}h)_{\mu\nu}=0$ and $(\mathcal{D}^{R}h)_{\mu\nu}=0$.
The third and fourth solutions are massive gravitons which are
solutions of $(\mathcal{D}^{m_1}h)_{\mu\nu}=0$ and
$(\mathcal{D}^{m_2}h)_{\mu\nu}=0$. To avoid any exponential
divergence at the boundary, the third and fourth solutions reduce to
the following solutions with the following parameter regions: $\\$
$\bullet$ $m^2> 0$ and $c_L\geq 0$ \bea
(h,\bar{h})&=&\Big(\frac{6-\frac{m^2\ell}{\mu}+\sqrt{2+4m^2\ell^2+\frac{m^4\ell^2}{\mu^2}}}{4},~~
\frac{-2-\frac{m^2\ell}{\mu}+\sqrt{2+4m^2\ell^2+\frac{m^4\ell^2}{\mu^2}}}{4}\Big),\nonumber\\
 (h,\bar{h})&=&\Big(\frac{-2+\frac{m^2\ell}{\mu}+
\sqrt{2+4m^2\ell^2+\frac{m^4\ell^2}{\mu^2}}}{4},~~\frac{6+\frac{m^2\ell}{\mu}+
\sqrt{2+4m^2\ell^2+\frac{m^4\ell^2}{\mu^2}}}{4}\Big);\nonumber\eea
\\$\bullet$ $\mu\ell\geq\frac{3}{4}$ and
$m^2\ell\leq-2\mu^2\ell-\mu\sqrt{4\mu^2\ell^2-2}$ and $c_L\geq0$\be
(h,\bar{h})=\Big(\frac{6-\frac{m^2\ell}{\mu}\pm
\sqrt{2+4m^2\ell^2+\frac{m^4\ell^2}{\mu^2}}}{4},~~\frac{-2-\frac{m^2\ell}{\mu}\pm
\sqrt{2+4m^2\ell^2+\frac{m^4\ell^2}{\mu^2}}}{4}\Big);\nonumber\ee
\\$\bullet$ Otherwise: we only have massless gravitons which do not blow up at the boundary.

For the mass of the graviton to be nonnegative, we also have to
impose the same conditions $\mu\ell\geq\frac{3}{4}$ and
$m^2\ell\leq-2\mu^2\ell-\mu\sqrt{4\mu^2\ell^2-2}$ or $m^2\geq 0$
along with the conditions $c_L\geq 0$. Thus in the remainder of this
paper, we only concentrate within this parameter region.


At the critical line $c_L=0$, the linearized equation of motion
defined using the mutually commuting operators (\ref{eee}) becomes
\be\label{log}(\mathcal{D}^{R}\mathcal{D}^{L}\mathcal{D}^{m_1}\mathcal{D}^{L}h)_{\mu\nu}=0,\ee
where $m_1=-\frac{1+2m^2\ell^2}{2\ell}$. From this expression we can
see that two of the operators degenerate here and the solutions
degenerate here, too. Thus we can construct the new ``log" solution
at $c_L=0$ following \cite{Grumiller:2008qz}. For $m^2> 0$, the left
moving massless and massive graviton solutions can be used to
construct a new log solution while for $\mu\ell\geq\frac{3}{4}$ and
$m^2\ell\leq-2\mu^2\ell-\mu\sqrt{4\mu^2\ell^2-2}$, the new log
solution can be obtained using the solution with the minus sign in
front of the square root. The log solutions constructed in these two
ways are the same and we have \be\label{newL}
\psi_{\mu\nu}^{newL}
=y(\tau,\rho)\psi^L_{\mu\nu},\ee where
$y(\tau,\rho)=-i\tau-\ln\cosh{\rho}$ and $\psi^L_{\mu\nu}$ is the
wavefunction for left moving massless mode. The new solution
satisfies
\be(\mathcal{D}^{L}\mathcal{D}^{L}h^{newL})_{\mu\nu}=0,~~~~(\mathcal{D}^{L}h^{newL})_{\mu\nu}\neq0.\ee
Note here that though this new solution was constructed within the
parameter regions shown above, the solution is a solution to
(\ref{log}) for arbitrary value of the mass parameters obeying
$c_L=0$.

At $c_L=0$, there is a special point
$m^2\ell^2=-2\mu\ell=-\frac{3}{2}$, at which $\mathcal{D}^{m_1}$
also becomes $\mathcal{D}^{L}$. The linearized equation of motion
becomes \be
(\mathcal{D}^{R}\mathcal{D}^{L}\mathcal{D}^{L}\mathcal{D}^{L}h)_{\mu\nu}=0.
\ee Thus the two solutions of massive gravitons both coincide with
the left moving massless solution. Remember that in the case of NMG
theory \cite{Liu:2009bk}, at the point $m^2\ell^2=1/2$ the two modes
of massive gravitons also become massless gravitons simultaneously.
However, in that theory, the two massive gravitons become one
left-moving and one right-moving massless modes. Here, because of a
non-zero $\mu$, parity is violated and the two massive modes can be
both reduced to two left-moving massless modes at a certain point.

Now because of this degeneration new solutions could appear which
obey
\be(\mathcal{D}^{L}\mathcal{D}^{L}\mathcal{D}^{L}h^{newS})_{\mu\nu}=0,~~~
(\mathcal{D}^{L}\mathcal{D}^{L}h^{newS})_{\mu\nu}\neq0,~~~~(\mathcal{D}^{L}h^{newS})_{\mu\nu}\neq0,\ee
and we find one such solution to be \be\label{newS}
\psi_{\mu\nu}^{newS}=Y(\tau,\rho)\psi^L_{\mu\nu},\ee where
$Y(\tau,\rho)=(-i\tau-\ln\cosh{\rho})^2$ and $\psi^L_{\mu\nu}$ is
the wavefunction of the left moving massless mode. This new solution
does not obey either Brown-Henneaux boundary conditions or log
boundary conditions. In order to accommodate this new solution we
will have to impose an even looser boundary condition than the log
boundary condition at this point. In the next section we will show
that at this special point an even looser boundary condition called
log-square boundary condition can be consistent.

\section{Consistent Boundary Conditions}
In this section, we will first derive the expression for conserved
charges and then study the consistency of the Brown-Henneaux and log
boundary conditions with this theory. We will also show that at the
special point $m^2\ell^2=-2\mu\ell=-\frac{3}{2}$ we can have another
consistent boundary condition.
\subsection{Conserved Charges}
In this subsection we will give the basic formulae to calculate the
conserved charges using the covariant formalism
\cite{{Barnich:2001jy},Barnich:2007bf,{Guica:2008mu},{Maloney:2009ck}}
(see also \cite{{abbottdeser},{iyerwald},{andersontorre},{torre},
{bbheneaux},{bbheneaux2},{barnichstokes},comperethesis})
 for this generalized massive gravity.

For convenience we define \be
{\cal{G}}_{\mu\nu}=R_{\mu\nu}-\frac{1}{2}g_{\mu\nu}R+\Lambda
g_{\mu\nu}.\ee
Then the covariant energy momentum tensor for the linearized
gravitional excitations of this new massive gravity theory can be
identified as \bea\label{energy} 32 \pi m^2G
T_{\mu\nu}&=&(2m^2+5\Lambda){\cal{G}}^{(1)}_{\mu\nu}-
\frac{1}{2}(\bar{g}_{\mu\nu}\bar{\nabla}^2-\bar{\nabla}_\mu\bar{\nabla}_\nu
+2\Lambda\bar{g}_{\mu\nu})R^{(1)}\nonumber\\
&&~-2(\bar{\nabla}^2{\cal{G}}^{(1)}_{\mu\nu}-\Lambda\bar{g}_{\mu\nu}
R^{(1)})+\frac{2m^2}{\mu}C^{(1)\mu\nu }.\eea It is shown in
\cite{{Deser:2002rt},Deser:2002jk} that when the background
spacetime admits a Killing vector $\xi_\mu$, the current
\be\label{tmunu} {\cal{K}}^\mu=  16\pi G\xi_\nu T^{\mu\nu}\ee is
covariantly conserved $\bar{\nabla}_{\mu}{\cal{K}}^\mu=0$. Then
there exists an antisymmetric two form tensor ${\cal{F}}^{\mu\nu}$
such that \be{\cal{K}}^\mu=16\pi
G\bar{\nabla}_{\nu}{\cal{F}}^{\mu\nu}\ee and the corresponding
charge could be written as a surface integral as \be
\label{conserve}Q(\xi)=-\frac{1}{8\pi
G}\int_{M}\sqrt{-\bar{g}}{\cal{K}}^0 =-\frac{1}{8\pi
G}\int_{\partial{M}}dS_i\sqrt{{-\bar{g}}}{\cal{F}}^{0i},\ee where
$\partial{M}$ is the boundary of a spacelike surface $M$. We have
chosen $M$ as constant time surface here and the expression is under
the coordinate system of (\ref{AdS3metric}).

We can rewrite each term in the expression of the energy momentum
tensor (\ref{energy}) as a total covariant derivative term
\cite{{Deser:2002rt},Deser:2002jk,{Deser:2003vh},{Bouchareb:2007yx}}
by using the definition and the properties of killing vectors: \be
\bar{\nabla}_{\mu}\xi^\mu=0, \hskip 0.5 cm
\bar{\nabla}_{\sigma}\bar{\nabla}^{\mu}\xi^\sigma=2\Lambda \xi^\mu,
\hskip 0.5 cm \bar{\nabla}^2\xi_\sigma=-2\Lambda \xi_\sigma\ee
 and the final result is \be  \xi_\nu T^{\mu\nu}={\bar{\nabla}}_{\sigma}
{\cal{F}}^{\mu\sigma},\ee
 where \bea
{\cal{F}}^{\mu\sigma}&=&(1-\frac{1}{2m^2\ell^2})F[\xi]^{\mu\sigma}+\frac{1}{\mu}F[\eta]^{\mu\sigma}
\nonumber\\&&-\frac{1}{4m^2}\Big\{\xi^\mu\bar{\nabla}^\sigma R^{(1)}
+R^{(1)}\bar{\nabla}^\mu\xi^\sigma-\xi^\sigma\bar{\nabla}^\mu
R^{(1)}\Big\}\nonumber \\
&&  -\frac{1}{m^2}\Big\{\xi_\nu \bar{\nabla}^{\sigma}
{\cal{G}}^{(1)\mu \nu} - \xi_\nu \bar{\nabla}^{\mu}
{\cal{G}}^{(1)\sigma \nu} - {\cal{G}}^{(1)\mu\nu}
\bar{\nabla}^{\sigma} \xi_\nu +
{\cal{G}}^{(1)\sigma\nu}\bar{\nabla}^{\mu}\xi_\nu\Big\}
\nonumber\\&&+\frac{1}{\mu}\Big\{\xi_\lambda(\varepsilon^{\mu\sigma\nu}{\cal{G}}^{(1)\lambda}_\nu
-\frac{1}{2}\varepsilon^{\mu\sigma\lambda}{\cal{G}}^{(1)})\Big\}\eea

and \bea F[\xi]^{\mu\sigma}&=&\frac{1}{2}\Big\{\xi_\nu
\bar{\nabla}^{\mu}h^{\sigma \nu} -\xi_\nu
\bar{\nabla}^{\sigma}h^{\mu\nu} +\xi^\mu \bar{\nabla}^\sigma h
-\xi^\sigma \bar{\nabla}^\mu h \nonumber \\
&&  + h^{\mu \nu}\bar{\nabla}^\sigma \xi_\nu - h^{\sigma
\nu}\bar{\nabla}^\mu \xi_\nu + \xi^\sigma \bar{\nabla}_{\nu}h^{\mu
\nu} -\xi^\mu \bar{\nabla}_{\nu}h^{\sigma \nu} + h\bar{\nabla}^\mu
\xi^\sigma\Big\},\nonumber\\
\eta_\mu&=&\frac{1}{2}\varepsilon_{\mu\nu\lambda}\bar{\nabla}^{\nu}\xi^{\lambda}.\nonumber\\
\eea

Thus the conserved charge (\ref{conserve}) becomes \be Q(\xi)
=-\frac{1}{8\pi
G}\int_{\partial{M}}dS_i\sqrt{{-\bar{g}}}{\cal{F}}^{0i}.\ee Here we
choose the spacelike surface as the constant time surface, then the
expression for the conserved charge could be simplified as
\be\label{concharge} Q(\xi) =-\lim_{\rho
\rightarrow\infty}\frac{1}{8\pi G}\int d\phi
\sqrt{{-\bar{g}}}{\cal{F}}^{0\rho},\ee where $\rho$ is the radial
coordinate of $AdS_3.$

Note that here to get the formula for the conserved charges, we have
used the definition of the killing vectors
$\bar{\nabla}_{\mu}\xi_{\nu}+\bar{\nabla}_{\nu}\xi_{\mu}=0$
 which does not hold any more for asymptotic
symmetries of the spacetime. Thus for asymptotic symmetries which do
not obey
$\bar{\nabla}_{\mu}\xi_{\nu}+\bar{\nabla}_{\nu}\xi_{\mu}=0$,
(\ref{tmunu}) is no longer a conserved quantity and we need to add
some terms composed by
$\bar{\nabla}_{\mu}\xi_{\nu}+\bar{\nabla}_{\nu}\xi_{\mu}$ and
$h_{\mu\nu}$ \cite{Compere:2008cv}. However, the formula
(\ref{concharge}) will still be valid to the linearized level of
gravitational excitations in our consideration.

\subsection{Brown-Henneaux Boundary Condition}
In this section we will analyze the Brown-Henneaux boundary
condition \cite{Brown:1986nw}  for the generalized massive gravity
theory. We will calculate the conserved charges corresponding to the
generators of the asymptotical symmetry under this boundary
condition and see whether all the charges are finite or not.  In
this and the following sections we will work in the global
coordinate system (\ref{AdS3metric}).

The Brown-Henneaux boundary condition for the linearized
gravitational excitations in asymptotical $AdS_3$ spacetime can be
written as
  \be
\left(
  \begin{array}{ccccc}
 h_{++}= {\mathcal{O}}(1) & h_{+-}= {\mathcal{O}}(1)  & h_{+\rho}= {\mathcal{O}}({e^{-2\rho}})  \\
 h_{-+}=h_{+-} & h_{--}= {\mathcal{O}}(1)  & h_{- \rho}= {\mathcal{O}}({e^{-2\rho}})  \\
  h_{\rho+}=h_{+\rho} &
  h_{\rho-}=h_{- \rho} & h_{\rho\rho}= {\mathcal{O}}({e^{-2\rho}}) \\
  \end{array}
\right) \ee in the global coordinate system.

The corresponding asymptotic Killing vectors are
\bea\label{bhkilling}  \xi &=& \xi^+\partial_++\xi^-\partial_-+\xi^{\rho}\partial_\rho \nonumber\\
&=&~~[\epsilon^+({\tau^+})
+2{e^{-2\rho}}\partial_-^2\epsilon^-({\tau^-})+
{\mathcal{O}}({e^{-4\rho}})]\partial_+
\nonumber\\
&&+~ [\epsilon^-({\tau^-})
+2e^{-2\rho}\partial_+^2\epsilon^+({\tau^+})+
{\mathcal{O}}(e^{-4\rho})]\partial_- \nonumber\\&& -\frac{1}{2}
[\partial_{+}{\epsilon^{+}({\tau^+})} +\partial_-\epsilon^-(\tau^-)+
{\mathcal{O}}({e^{-2\rho}})]\partial_\rho. \eea

Because $\phi$ is periodic, we could choose the basis
$\epsilon_m^+=e^{im\tau^+}$ and $\epsilon_n^-=e^{in\tau^-}$ and
denote the corresponding killing vectors as $\xi^L_m$ and $\xi^R_n$
. The algebra structure of these vectors is \be\label{virasoro}
i[\xi^L_m, \xi^L_n]=(m-n)\xi^L_{m+n},~~i[\xi^R_m,
\xi^R_n]=(m-n)\xi^R_{m+n},~~[\xi^L_m, \xi^R_n]=0.\ee
 Thus these asymptotic Killing vectors give two copies of Virasora
algebra. To calculate the conserved charges using (\ref{concharge})
we first parameterize the gravitons as follows
\bea\label{bhcon} h_{++}&=&f_{++}(\tau,\phi)+\dots\nonumber\\
h_{+-}&=&f_{+-}(\tau,\phi)+\dots\nonumber\\
h_{+\rho}&=&e^{-2\rho}f_{+\rho}(\tau,\phi)+\dots\nonumber\\
h_{--}&=&f_{--}(\tau,\phi)+\dots\nonumber\\
h_{-\rho}&=&e^{-2\rho}f_{-\rho}(\tau,\phi)+\dots\nonumber\\
h_{\rho\rho}&=&e^{-2\rho}f_{\rho\rho}(\tau,\phi)+\dots.,\eea where
$f_{\mu\nu}$ depends only on $\tau$ and $\phi$ while not on $\rho$
and the ``$\dots$" terms are lower order terms which do not
contribute to the conserved charges. After plugging (\ref{bhcon})
into (\ref{concharge}) and performing the $\rho \rightarrow\infty$
limit, we obtain \bea\label{bhcharge} Q&=&\frac{1}{8\pi G\ell}\int
d\phi
\Big\{(1-\frac{1}{2m^2\ell^2}-\frac{1}{\mu\ell})\epsilon^+f_{++}
+(1-\frac{1}{2m^2\ell^2}+\frac{1}{\mu\ell})\epsilon^-f_{--}\nonumber\\
&&~~~~~~-(1+\frac{1}{2m^2\ell^2})\frac{(\epsilon^++\epsilon^-)(16f_{+-}-f_{\rho\rho})}{16}\Big\}\eea
for this theory. Three components of the equation of motion
(\ref{eomold}), which do not involve second derivative terms, can be
viewed as asymptotic constraints. The $\rho\rho$ component gives
 \be16f_{+-}-f_{\rho\rho}=0\ee at the boundary and the $+\rho$ and
 $-\rho$ components give \be
(1-\frac{1}{2m^2\ell^2}-\frac{1}{\mu\ell})\partial_-f_{++}=(1-\frac{1}{2m^2\ell^2}+\frac{1}{\mu\ell})\partial_+f_{--}=0\ee
 respectively. After plugging in these boundary constraints, the
 conserved charges become
\bea Q&=&\frac{1}{8\pi G\ell}\int d\phi
\Big\{(1-\frac{1}{2m^2\ell^2}-\frac{1}{\mu\ell})\epsilon^+f_{++}
+(1-\frac{1}{2m^2\ell^2}+\frac{1}{\mu\ell})\epsilon^-f_{--}\Big\}\nonumber\\~\nonumber\\
&=&Q_{L}+Q_{R},\eea where the left moving conserved charge is \be
Q_{L}=\frac{1}{8\pi G\ell}\int d\phi
\big\{(1-\frac{1}{2m^2\ell^2}-\frac{1}{\mu\ell})\epsilon^+f_{++}\big\},\ee
and the right moving conserved charge \be Q_{R}=\frac{1}{8\pi
G\ell}\int d\phi
\big\{(1-\frac{1}{2m^2\ell^2}+\frac{1}{\mu\ell})\epsilon^-f_{--}\big\}.\ee
The left moving and right moving conserved charges fulfill two
copies of Virasoro algebra with central charges \be
c_L=\frac{3\ell}{2G}(1-\frac{1}{2m^2\ell^2}-\frac{1}{\mu \ell}),~~~~
c_R=\frac{3\ell}{2G}(1-\frac{1}{2m^2\ell^2}+\frac{1}{\mu \ell}). \ee
We can see that the conserved charges $Q$ are always finite for
arbitrary value of $m,~\mu$, so the Brown-Henneaux boundary
condition is always consistent with the generalized massive gravity
theory. At the critical line $\frac{1}{2m^2\ell^2}+\frac{1}{\mu
\ell}=1$, the left moving conserved charges vanish which shows that
the generalized massive gravity theory is chiral at the critical
line under the Brown-Henneaux boundary condition. We can see that
the TMG chiral gravity at $\mu\ell=1$ is just the special case
$m^2\rightarrow\infty$ of this generalized gravity theory.

\subsection{Log Boundary Condition}
As we have checked in the previous section, at the critical line
$\frac{1}{2m^2\ell^2}+\frac{1}{\mu \ell}=1$, new solutions of the
equation of motion (\ref{eomold}) appeared. The new solution does
not obey the Brown-Henneaux boundary conditions. In order to include
these new interesting solutions, the boundary conditions need to be
loosened \cite{{Grumiller:2008es},{Henneaux:2009pw},Maloney:2009ck}.
Earlier investigations on the relaxation of the boundary conditions
for gravity coupled with scalar fields in anti-de Sitter spacetime
could be found in
\cite{{Henneaux:2002wm},{Henneaux:2004zi},{Hertog:2004dr},{Henneaux:2006hk},Amsel:2006uf}.

We relax the boundary condition as follows to include the solution
$\psi^{\rm newL}_{\mu\nu}$:
  \be
\left(
  \begin{array}{ccccc}
 h_{++}= {\mathcal{O}}(\rho) & h_{+-}= {\mathcal{O}}(1)  & h_{+\rho}= {\mathcal{O}}({
\rho e^{-2\rho}})  \\
 h_{-+}=h_{+-} & h_{--}= {\mathcal{O}}(1)  & h_{- \rho}= {\mathcal{O}}({e^{-2\rho}})  \\
  h_{\rho+}=h_{+\rho} &
  h_{\rho-}=h_{- \rho} & h_{\rho\rho}= {\mathcal{O}}({e^{-2\rho}}) \\
  \end{array}
\right). \ee Then the corresponding asymptotic Killing vector can be
calculated to be
\bea  \xi &=& \xi^+\partial_++\xi^-\partial_-+\xi^{\rho}\partial_\rho \nonumber\\
&=&~~[\epsilon^+({\tau^+})
+2{e^{-2\rho}}\partial_-^2\epsilon^-({\tau^-})+
{\mathcal{O}}({e^{-4\rho}})]\partial_+
\nonumber\\
&&+~ [\epsilon^-({\tau^-})
+2e^{-2\rho}\partial_+^2\epsilon^+({\tau^+})+ {\mathcal{O}}(\rho
e^{-4\rho})]\partial_- \nonumber\\&& -\frac{1}{2}
[\partial_{+}{\epsilon^{+}({\tau^+})} +\partial_-\epsilon^-(\tau^-)+
{\mathcal{O}}({e^{-2\rho}})]\partial_\rho. \eea Note that these
asymptotic Killing vectors are different from (\ref{bhkilling}) only
in the subleading order, so these also give two copies of Varasoro
algebra the same as (\ref{virasoro}).

With this new boundary condition we can parameterize the asymptotic
excitations as follows
\bea\label{logleft} h_{++}&=&\rho f^{L}_{++}(\tau,\phi)+\dots\nonumber\\
h_{+-}&=&
f^{L}_{+-}(\tau,\phi)+\dots\nonumber\\
h_{+\rho}&=&\rho e^{-2\rho}
f^{L}_{+\rho}(\tau,\phi)+\dots\nonumber\\
h_{--}&=&f^{L}_{--}(\tau,\phi)+\dots\nonumber\\
h_{-\rho}&=&e^{-2\rho}f^{L}_{-\rho}(\tau,\phi)+\dots\nonumber\\
h_{\rho\rho}&=&e^{-2\rho}f^{L}_{\rho\rho}(\tau,\phi)+\dots .\eea
Note that $f^{L}_{\mu\nu}$ depends only on $\tau$, $\phi$ while not
on $\rho$ and the ``$\dots$" terms are subleading terms which do not
contribute to the conserved charge. After plugging (\ref{logleft})
into (\ref{concharge}) and performing the $\rho \rightarrow\infty$
limit, we could obtain \bea Q&=&\frac{1}{8\pi G\ell}\int d\phi
\Big\{(1-\frac{1}{2m^2\ell^2}-\frac{1}{\mu\ell})\epsilon^+f^{L}_{++}\cdot\infty+
(\frac{2}{m^2\ell^2}+\frac{1}{\mu\ell})\epsilon^+f^{L}_{++}\nonumber\\
&&~~~+(1-\frac{1}{2m^2\ell^2}+\frac{1}{\mu\ell})\epsilon^-f^{L}_{--}
-(1+\frac{1}{2m^2\ell^2})\frac{(\epsilon^++\epsilon^-)(16f^{L}_{+-}-f^{L}_{\rho\rho})}{16}\Big\}.\eea
The first term is a linear divergent term proportional to $\rho$ at
infinity, which is caused by the relaxation of the boundary
condition. We see that the conserved charges can only be finite at
the critical line $\frac{1}{2m^2\ell^2}+\frac{1}{\mu \ell}=1$, which
means that the log boundary condition is only well-defined at the
line $\frac{1}{2m^2\ell^2}+\frac{1}{\mu \ell}=1$. The asymptotic
constraints coming from the equation of motion (\ref{eomold}), which
are \be 16f^{L}_{+-}-f^{L}_{\rho\rho}=0,\ee and \be
(\frac{2}{m^2\ell^2}+\frac{1}{\mu\ell})\partial_- f^{L}_{++}=
\partial_+ f^{L}_{--}=0.\ee Now at the critical line
$\frac{1}{2m^2\ell^2}+\frac{1}{\mu \ell}=1$, the conserved charges
become \be Q_L=\frac{1}{8\pi G\ell}\int d\phi
\big[(1+\frac{3}{2m^2\ell^2})\epsilon^+f^{L}_{++}\big], ~~~~
Q_{R}=\frac{1}{4\pi G\ell}\int d\phi
\big[\frac{1}{\mu\ell}\epsilon^-f^{L}_{--}\big]. \ee Thus we have
loosened the boundary condition to get nonzero left moving charges.
The central charges are not altered with the same argument for TMG
in \cite{Maloney:2009ck}.

Note that at the special point $m^2\ell^2=-2\mu\ell=-\frac{3}{2}$,
$Q_L=0$ under this boundary condition which means that the theory is
still chiral and there is the possibility that we can further relax
the boundary condition to have nonzero left moving conserved
charges. This coincides with the fact that at this special point new
solutions which do not obey the log boundary conditions emerge. In
the next subsection we will give the new boundary condition at this
special point.

\subsection{Log-Square Boundary Condition} At the critical point
$m^2\ell^2=-2\mu\ell=-\frac{3}{2}$, a kind of new solution
(\ref{newS}) appeared. This new solution does not obey either the
Brown-Henneaux boundary conditions or the log boundary condition,
and we have to relax the boundary condition again to accommodate
such kind of solutions.

With the hint from the asymptotical behavior of the new solution
(\ref{newS}) we can relax the boundary condition as follows to
include the solution $\psi^{\rm newS}_{\mu\nu}$
  \be
\left(
  \begin{array}{ccccc}
 h_{++}= {\mathcal{O}}(\rho^2) & h_{+-}= {\mathcal{O}}(1)  & h_{+\rho}= {\mathcal{O}}({
\rho^2 e^{-2\rho}})  \\
 h_{-+}=h_{+-} & h_{--}= {\mathcal{O}}(1)  & h_{- \rho}= {\mathcal{O}}({e^{-2\rho}})  \\
  h_{\rho+}=h_{+\rho} &
  h_{\rho-}=h_{- \rho} & h_{\rho\rho}= {\mathcal{O}}({e^{-2\rho}}) \\
  \end{array}
\right). \ee Then the corresponding asymptotic Killing vector can be
calculated to be
\bea  \xi &=& \xi^+\partial_++\xi^-\partial_-+\xi^{\rho}\partial_\rho \nonumber\\
&=&~~[\epsilon^+({\tau^+})
+2{e^{-2\rho}}\partial_-^2\epsilon^-({\tau^-})+
{\mathcal{O}}({e^{-4\rho}})]\partial_+
\nonumber\\
&&+~ [\epsilon^-({\tau^-})
+2e^{-2\rho}\partial_+^2\epsilon^+({\tau^+})+ {\mathcal{O}}(\rho^2
e^{-4\rho})]\partial_- \nonumber\\&& -\frac{1}{2}
[\partial_{+}{\epsilon^{+}({\tau^+})} +\partial_-\epsilon^-(\tau^-)+
{\mathcal{O}}({e^{-2\rho}})]\partial_\rho. \eea Note that these
asymptotic Killing vectors are different from (\ref{bhkilling}) only
in the subleading order, so these also give two copies of Varasoro
algebra the same as (\ref{virasoro}).

With this new boundary condition we can parameterize the asymptotic
excitations as follows
\bea\label{logsquare} h_{++}&=&\rho^2 f^{S}_{++}(\tau,\phi)+\dots\nonumber\\
h_{+-}&=&
f^{S}_{+-}(\tau,\phi)+\dots\nonumber\\
h_{+\rho}&=&\rho^2 e^{-2\rho}
f^{S}_{+\rho}(\tau,\phi)+\dots\nonumber\\
h_{--}&=&f^{S}_{--}(\tau,\phi)+\dots\nonumber\\
h_{-\rho}&=&e^{-2\rho}f^{S}_{-\rho}(\tau,\phi)+\dots\nonumber\\
h_{\rho\rho}&=&e^{-2\rho}f^{S}_{\rho\rho}(\tau,\phi)+\dots.\eea

Note that $f^{S}_{\mu\nu}$ depends only on $\tau$, $\phi$ while not
on $\rho$ and the ``$\dots$" terms are subleading terms which do not
contribute to the conserved charge. After plugging (\ref{logsquare})
into (\ref{concharge}) and performing the $\rho \rightarrow\infty$
limit, we could obtain \bea Q&=&\frac{1}{8\pi G\ell}\int d\phi
\Big\{(1-\frac{1}{2m^2\ell^2}-\frac{1}{\mu\ell})\epsilon^+f^{S}_{++}\cdot[\lim_{\rho\rightarrow\infty}\rho^2]
-(1-\frac{9}{2m^2\ell^2}-\frac{3}{\mu\ell})\epsilon^+f^{S}_{++}\cdot[\lim_{\rho\rightarrow\infty}\rho]\nonumber\\
&&~~~~~-(\frac{4}{m^2\ell^2}+\frac{1}{\mu\ell})\epsilon^+f^{S}_{++}
+(1-\frac{1}{2m^2\ell^2}+\frac{1}{\mu\ell})\epsilon^-f^{S}_{--}\nonumber\\
&&~~~~~~-(1+\frac{1}{2m^2\ell^2})\frac{(\epsilon^++\epsilon^-)(16f^{S}_{+-}-f^{S}_{\rho\rho})}{16}\Big\}.\eea
Note that we did not include the next to leading order term in the
asymptotic behavior (\ref{logsquare}), i.e. the term proportional to
$\rho$ in $h_{++}$. This does not affect our result because the
conserved charges are linear in $h_{\mu\nu}$. We see that the
conserved charges can only be finite at the critical point
$m^2\ell^2=-2\mu\ell=-\frac{3}{2}$, which means that this log-square
boundary condition is only well-defined at the special point
$m^2\ell^2=-2\mu\ell=-\frac{3}{2}$. The asymptotic constraints
coming from the equation of motion (\ref{eomold}) are \be
16f^{S}_{+-}-f^{S}_{\rho\rho}=0,\ee and\be
\partial_- f^{S}_{++}=\partial_+ f^{S}_{--}=0\ee Now at the point $m^2\ell^2=-2\mu\ell=-\frac{3}{2}$,
the conserved charges become \be Q_L=\frac{1}{6\pi G\ell}\int d\phi
\big[\epsilon^+f^{S}_{++}\big], ~~~~ Q_{R}=\frac{1}{3\pi G\ell}\int
d\phi \big[\epsilon^-f^{S}_{--}\big].\ee The central charges are not
changed because the Lie derivative of $f^S_{++}$ also involves no
terms like $\partial_+\epsilon^++ \partial_+^3\epsilon^+$ which
contributes to the central charge.

\section{Conclusion and Discussion}
In this note we have studied the generalized massive gravity in
asymptotically $AdS_3$ spacetime by combining the two mass terms of
TMG and NMG. This generalized massive gravity no longer possesses
the left-right symmetry because of the addition of the Chern-Simons
term. Thus in this theory there exists a chiral line $c_L=0$ at
which the theory becomes chiral and only right moving modes exist.
The conserved charges are calculated to show that the left moving
conserved charges indeed vanish at $c_L=0$ under Brown-Henneaux
boundary condition.

At the linearized level, we calculated the highest weight graviton
solutions of this massive gravity. Very similar to the case of TMG,
at $c_L=0$ one massive graviton coincides with the left moving
massless mode and a new solution with log asymptotic behavior
appeared. Thus at the critical line $c_L=0$ the boundary condition
can be loosened to the log boundary condition and we showed that
this boundary condition is indeed consistent with the theory at
$c_L=0$. However, because we have two adjustable mass parameters in
this generalized massive gravity, a new special point emerged which
does not exist in TMG or NMG. At this special point, both of the
massive graviton solutions become left moving massless modes and we
can have another new solution besides the one which also exists in
TMG and NMG. This new solution has a log-square asymptotic behavior
and we showed that a new boundary condition called log-square
boundary condition which can accommodate this new solution can be
consistent with the theory at this special point.

It has been suggested in
\cite{Grumiller:2008qz,{Sachs:2008yi},{Myung:2008dm}} that the
logarithmic CFT may be the dual field theory of TMG at the chiral
point with the log boundary condition. It would be a challenge if we
can find the dual field theory which has similar properties with
this GMG at the special point where the log-square boundary
condition can be imposed.

It is interesting to find whether there are classical solutions that
exhibit the log square asymptotic behavior, like the solutions in
\cite{AyonBeato:2004fq,{AyonBeato:2005qq},{Carlip:2008eq},{Gibbons:2008vi},{Garbarz:2008qn}}
for TMG. Also it would be interesting to study other consistent
boundary conditions for this generalized massive gravity in
asymptotically $AdS_3$ at both the critical line and other value of
the mass parameters. To gain further understanding of the dual field
theory of this generalized massive gravity, we need to study the
classical solutions of this theory, which contribute to the sum over
geometry. The entropy of the BTZ black hole calculated using the
central charges obtained under Brown-Henneaux or log (square)
boundary conditions is the same to the one calculated using other
methods, e.g. in \cite{Sahoo:2006vz}. This is also the case for NMG
\cite{{Clement:2009gq},Liu:2009bk,{Liu:2009kc}}. We should further
study this subject to see if there can be other black hole solutions
under these boundary conditions.

\section*{Acknowledgments}

We would like to thank Prof. Rong-Gen Cai for useful discussions and
reading the manuscript. We would also like to thank Prof. Tianjun Li
for encouragements. This work was supported in part by the Chinese
Academy of Sciences with Grant No. KJCX3-SYW-N2 and the NSFC with
Grant No. 10821504 and No. 10525060.

\end{document}